\documentclass [12pt]{aastex}
\usepackage{graphics,graphicx}

\begin{document}

\title{Infrared Space Observatory 
Polarimetric Imaging of \\ the Egg Nebula (RAFGL~2688)\thanks{Based on
observations with ISO (Kessler et al. 1996), an ESA project with
instruments funded by ESA Member States (especially the PI countries:
France, Germany, the Netherlands and the United Kingdom) with the
participation of ISAS and NASA.}}

\author{
Joel H. Kastner\thanks{Chester F. Carlson Center
for Imaging Science, Rochester Institute  
of Technology, 54 Lomb Memorial Dr., Rochester, NY 14623},
Jingqiang Li$^{2}$, 
Ralf Siebenmorgen\thanks{European Southern Observatory,
Karl-Schwarzschildstr.\ 2, 85748 Garching b.\ Munich, Germany}
and David A. Weintraub\thanks{Dept.\ of Physics and
Astronomy, Vanderbilt University, 
Nashville, TN, 37235, U.S.A.}
} 

\begin{abstract}
We present polarimetric imaging of the protoplanetary nebula
RAFGL~2688 obtained at 4.5 $\mu$m with the Infrared Space
Observatory (ISO). We have deconvolved the images to remove
the signature of the point spread function of the ISO
telescope, to the extent possible. The deconvolved 4.5 $\mu$m
image and polarimetric map reveal a bright point source with faint,
surrounding reflection nebulosity. The reflection nebula is brightest to the
north-northeast, in agreement with previous ground- and
space-based infrared imaging. Comparison with previous near-infrared 
polarimetric imaging suggests that the polarization of
starlight induced by the dust grains in RAFGL~2688 is more or
less independent of wavelength between 2 $\mu$m and 4.5
$\mu$m. This, in turn, indicates that 
scattering dominates over thermal emission at wavelengths as
long as $\sim5$ $\mu$m, and that the dust grains have
characteristic radii $< 1$ $\mu$m. 
\end{abstract}
\keywords{stars: mass loss --- protoplanetary nebulae:
individual (AFGL 2688) -- dust --- polarization}

\received{}
\accepted{}

\section{Introduction}

Stellar mass loss
during post-main sequence stellar evolution is, more often than not,
an axisymmetric, rather than spherically symmetric, process. However, 
a coherent
explanation for the transformation from quasi-spherical red giant to 
bipolar planetary nebula remains elusive. During pre-main sequence
evolution, bipolar outflow is evidently an inevitable consequence of the
presence of a circumstellar, protoplanetary disk. But --- neglecting
the possibility that some evolved stars 
may retain massive, long-lived, ``fossil'' protoplanetary disks 
(Kastner \& Weintraub 1995) --- there is no obvious
natural ``collimating agent'' during post-main sequence stellar evolution.
Hence the shaping of bipolar planetary nebulae (PNs), the endpoints of 
stellar evolution for intermediate-mass (1-8 $M_\odot$) stars, 
is a topic of considerable contemporary interest in astronomy (see reviews
in Kastner, Soker, \& Rappaport 2000a).  

The object RAFGL~2688 (The Egg Nebula) has long been
regarded as prototypical of the class of evolved bipolar 
nebulae. The nebula was first identified as a bright far-IR source 
associated with axisymmetric optical nebulosity
by Ney et al.\ (1975). Based on its radio molecular emission
line profiles, Zuckerman et  al.\ (1976)
proposed that the object was representative of the rapid
transition from red giant to planetary nebula. 
Immediately following its discovery,  RAFGL~2688 
was established as a source of 
highly polarized optical and near-IR emission (e.g., Jones \& Dyck 1978).
The large net polarization in these early, large aperture measurements
was thought to originate with the 
scattering of starlight by circumstellar dust, where the
dust grain population is distributed in axisymmetric
fashion about the central star. The latitudinal dependence of the dust
density results in large opacities along the equatorial plane but 
relatively small opacities toward higher latitudes. Assuming the polar axis 
is nearly in the plane of the sky, the star 
itself would then be obscured along our line of sight, whereas starlight
can still escape along polar (but not equatorial) directions
and be scattered toward Earth. 

This qualitative model of the structure of RAFGL~2688 has since
been confirmed, first by Monte Carlo scattering models
(Yusef-Zadeh et al.\ 1984) and, most recently, by high
spatial resolution Hubble
Space Telescope near-infrared polarimetric imaging observations (Sahai
et al.\ 1998a; Fig.\ 1). The HST imaging demonstrates conclusively
that the nebula displays the centrosymmetric patterns and large degree
of linear polarization expected for an axisymmetric distribution of
dust grains illuminated by a central source. Furthermore, the HST
observations have established the position of the central illuminator, 
relative to the extended nebulosity, to within $\sim100$ AU (Weintraub
et al.\ 2000). In addition to the unseen illuminator, RAFGL~2688
contains a prominent, central source of direct (unpolarized) 
emission that is only detectable at wavelengths $>1.5 \mu$m (Sahai
et al.\ 1998a; Weintraub et al.\ 2000). The
relationship of this source to the illuminator is uncertain,
although the simplest explanation would be that the pair constitutes
a widely separated binary. 

Although HST has well established the optical 
and near-infrared morphology of RAFGL~2688
(Sahai et  al.\ 1998a,b), a detailed understanding of 
the nebula remains elusive. The kinematics of its 
molecular emission are particularly puzzling
(see Cox et al.\ 2000, Kastner et al.\ 2001,  and references
therein). This emission seems 
to trace multipolar, radially directed outflows, some
with no counterparts at optical or infrared wavelengths; alternatively,
the emission may be interpreted in terms of a combination of radial and
azimuthal velocity fields.
Deepening the mystery of the structure of RAFGL~2688 are recent
mid-infrared and radio imaging observations of dust emission that appear
to trace highly collimated (and perhaps precessing) outflows
(Morris \& Sahai 2000; Jura et al.\ 2000). 

Polarimetric imaging observations remain central to an understanding
of the structure and evolution of RAFGL~2688. While the nebula is well
characterized polarimetrically in the optical and near-infrared,
little or no spatially resolved polarimetric imaging data exists for
wavelengths $>2 \mu$m.  Here, we report on infrared polarimetric
imaging of RAFGL~2688 at 4.5 $\mu$m with the Infrared Space
Observatory (ISO, Kessler et al. 1996). In conjunction with previous
HST near-infrared polarimetric imaging and ground-based direct
mid-infrared imaging, the ISO polarimetric imaging yields insights
into the relative contributions of direct and unpolarized starlight
and of the grain size distribution in RAFGL~2688.

\section{Observations}

Data presented here were obtained with the ISO direct imaging
instrument, ISOCAM (Cesarsky et al., 1996), using its long wavelength
(LW) camera and passband filter LW1. The detector format was 32$\times$32
pixels. The response of LW1 is centered at 4.5 $\mu$m and is 
relatively flat, extending from 4 to 5\,$\mu$m.

ISOCAM was equipped with polarizers oriented at 0, 120, and 240
degrees.  Polarimetric imaging was set up according to the standard
observing template CAM05 (Siebenmorgen~1996). The corresponding
ISO observation numbers, TDT, are: 76800602, 76800703, 76800804,
76800905 76801006,76801107, 76801208, 76801309, 76801410 and 76801511.
We used a $4 \times 4$ raster with a raster step size of $6''$,
the $1.5''$ lens, a readout time for
each exposure of 2.1 s, and a gain of
one. Initially, we took 155 exposures on the source to
stabilize the detector and then a raster through the free ``hole'' of
the entrance wheel.  At each raster position 26 exposures were read
out.  This procedure was carried out for each of the three polarizers.
The polarizer rasters were repeated in three observing cycles.

\section{Results}

\subsection{Processed ISOCAM LW1 images}

The data were reduced with the ISOCAM interactive analysis
system (CIA version 4.0, Ott et al.\ 1996).  Only basic
reduction steps, such as dark current subtraction, removal
of cosmic ray hits and transient correction, are applied to
the data.  These steps are described together with
calibration uncertainties in Blommaert et al.~(2001).  The
coadded images at each raster position were projected on the
sky to derive the final mosaics.  The mosaics have a $66''
\times 68''$ total field of view and a pixel scale of
1.5$''$.  The polarized signal is found to be consistent
between the individual observing cycles. The average of all
cycles gives the final mosaic image for each polarizer.  The
output of this image processing procedure therefore
consisted of four images, each of total integration time 1124\,s:
three polarized images (one for each of the three
polarization filters) and one ``unpolarized'' or total
intensity image (obtained through the ``hole'' position).

As part of the ISO calibration program, the zodiacal light
and a set of (unpolarised) photometric standard stars was observed by
repeating CAM05 observing templates. From those calibration
measurements an instrumental polarization of $<$1.5\% is derived
(Siebenmorgen 1999). This instrumental polarization is small
enough to have negligible effect on the results described in
this paper. 

\begin{figure}[htb]
\includegraphics[scale=0.9,angle=90]{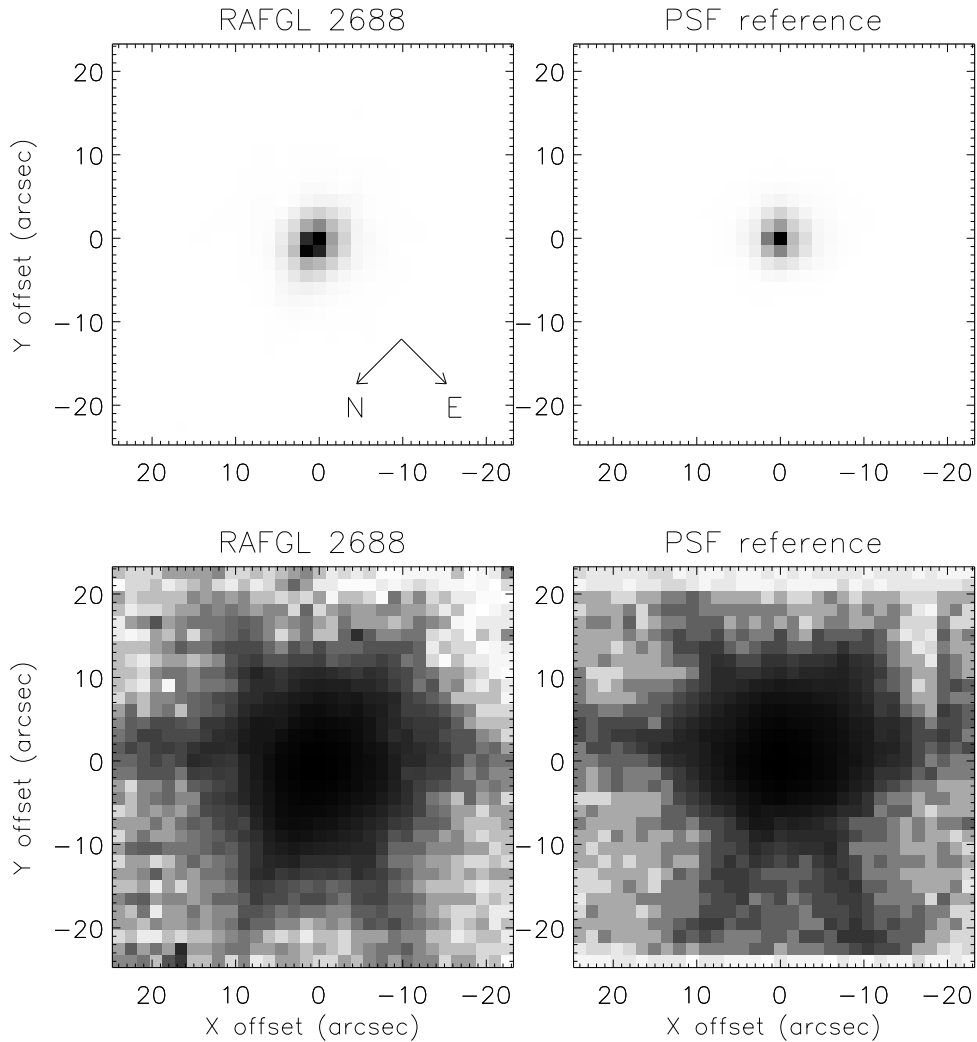}
\caption{ ISOCAM LW1 (4.5 $\mu$m) total 
intensity images of RAFGL~2688 
(left panels) and the standard star HIC~85317 (right panels). Images
at top are displayed with a linear greyscale representation,
while images at bottom are displayed with a
nonlinear greyscale emphasizing low-level structure
in the ISOCAM LW1 point spread function. Images are displayed in the
orientation in which they were obtained by the detector, i.e., with image
rows and columns representing detector rows and columns, respectively. }
\label{fig:hole_PSF}
\end{figure}

\begin{figure}[htb]
\includegraphics[scale=0.9,angle=90]{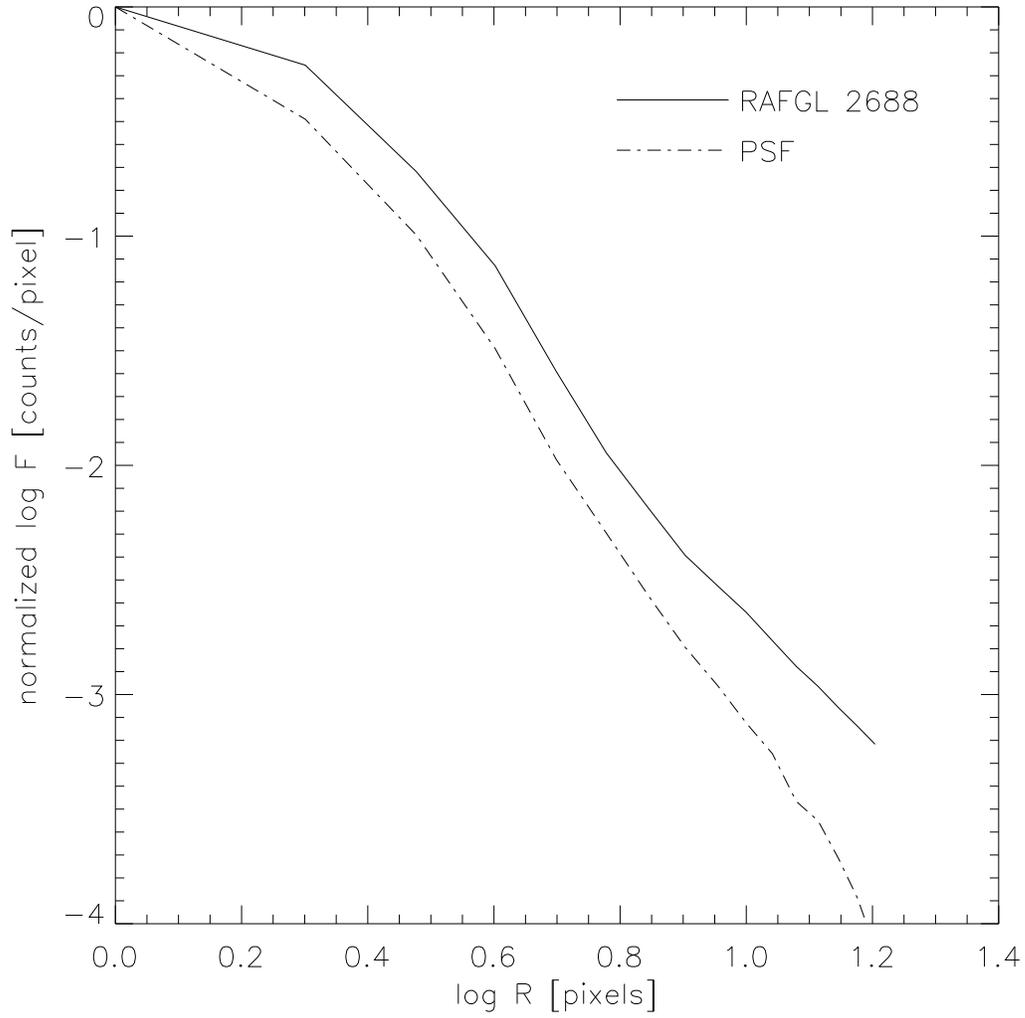}
\caption{ Profiles of 
intensity vs. radius for RAFGL
2688 (solid curve) and standard star (dashed curve). Both curves are
normalized to their peak intensities.}
\label{fig:radial_profiles}
\end{figure}

The total intensity image of RAFGL~2688 is displayed in Fig.\
\ref{fig:hole_PSF}, alongside a 3420\,s ISOCAM LW1 image of the
infrared-bright standard star HIC~85317 (spectral type A4V,
V$=6.5$mag, TDT=35600401). The comparison emphasizes the rather
complex, asymmetric point spread function (PSF) of ISOCAM at
4.5\,$\mu$m. The six-pointed structure of the PSF is due to
the support structure for the secondary mirror of the
telescope. 
Although
this point spread function dominates the structure of both the RAFGL
2688 and standard star images, it is also apparent that RAFGL~2688 is
extended relative to the standard. This is seen most clearly in the
comparisons of nonlinear greyscale images (Fig.\ \ref{fig:hole_PSF},
bottom panels) and plots of intensity vs.\ radius for RAFGL~2688 and
the standard star (Fig.\ \ref{fig:radial_profiles}), 
the latter of which reveals twice as much or more light in
the RAFGL 2688 images as in the PSF at distances $\sim10''$ from the
bright central PSF core.

\subsection{ISOCAM 4.5 $\mu$m polarimetric map}
\label{sec:rawpolmap}

The polarimetric image set obtained with the three polarizing 
filters in the LW camera is sufficient to uniquely determine
the Stokes parameters for each image pixel $(x,y)$. Specifically, we find
$I(x,y), Q(x,y), U(x,y)$ from the image set 
$P_{0}(x,y), P_{120}(x,y), P_{240}(x,y)$ as follows:
\begin{eqnarray}
I(x,y) = \frac{2}{3} [P_{0}(x,y) + P_{120}(x,y) +
P_{240}(x,y)] \\
Q(x,y) = \frac{2}{3} [2 P_{0}(x,y) - P_{120}(x,y) -
P_{240}(x,y)] \\
U(x,y) = -\frac{2}{\sqrt{3}}P_{120}(x,y) 
+ \frac{2}{\sqrt{3}}P_{240}(x,y)
\end{eqnarray}
The degree $p$ and position angle $\theta$ of linear polarization 
may then be derived from the Stokes parameters in the usual manner,
i.e., 
\begin{equation}
p(x,y) = \frac{(Q^2(x,y) + U^2(x,y))^{1/2}}{I(x,y)}
\end{equation}
and
\begin{equation}
\theta(x,y) = \frac{1}{2} \tan^{-1}\frac{U(x,y)}{Q(x,y)}.
\end{equation}
Polarization maps so obtained from ISOCAM polarimetric
imaging have been presented for the Crab nebula (Tuffs
et al.\ 1999), the protostellar system HH~108MMS (Siebenmorgen \& Kr\"ugel
2001), the starburst NGC~1808 galaxy (Siebenmorgen et al., 2001), and a
sample of ultraluminous galaxies (Siebenmorgen \& Efstathiou
2001).

\begin{figure}[htb]
\includegraphics[scale=0.9,angle=90]{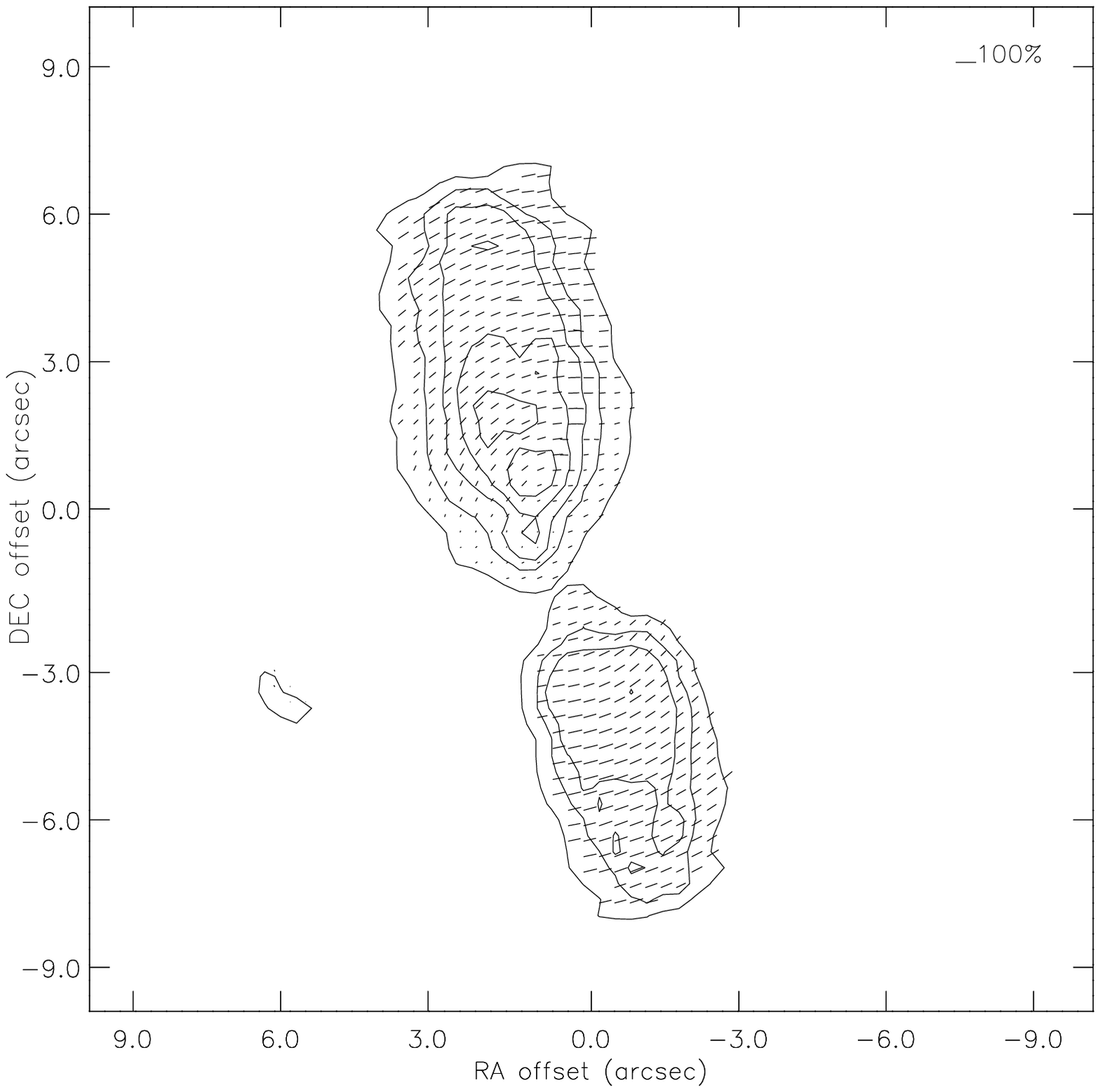}
\caption{ Hubble Space Telescope near-infrared (2 $\mu$m)
image (contours) and polarimetric map obtained with the Near-Infrared Camera
and Multi-Object Spectrometer (NICMOS), constructed
from archival images. Vectors indicate the degree and
position angle of polarization at each position across the image. See also
Sahai et al.\ (1998a) and Weintraub et al.\ (2000).} 
\label{fig:NICpol}
\end{figure}

The ISO polarization map obtained for RAFGL~2688 (Fig.\
\ref{fig:raw_polmap}) shows hints of centrosymmetry,
particularly in its northern regions.  The 
center of symmetry of this portion of the polarimetric map appears to
lie near the bright central region of the image.  This spatial
polarization structure suggests that much of the northern region of
the ISO 4.5 $\mu$m image of RAFGL~2688 is, in fact, scattered light,
and that the source of illumination lies in the vicinity of the image
intensity peak. However, the structure of the ISO polarimetric map is,
overall, much less well organized than that of the HST/NICMOS
polarization map (Fig.\ \ref{fig:NICpol}). In particular, 
the relative contributions from nebulosity and PSF are not
clear from Fig.\ \ref{fig:NICpol}; nor is it obvious
whether any extended 4.5 $\mu$m
emission is dominated by scattering (as is undoubtedly the case at 2
$\mu$m) or by direct, thermal emission. Evidently, either the ISO 
polarimetric map suffers from systematic distortions, or there are
significant differences in the spatial distribution and
degree of polarization in RAFGL~2688 between 2 $\mu$m and 4.5 $\mu$m.

\section{Reconstruction of Direct and Polarimetric Images}


To investigate the extent to which the prominent signature of the PSF in the 
individual polarimetric images affects the ISO 4.5 $\mu$m polarimetric 
map, we attempted to remove the PSF signature from these images
and that of the total intensity image. To this end we 
have employed two popular deconvolution methods, maximum 
likelihood (hereafter ML; Richardson 1972; Lucy 1974) 
and maximum entropy (e.g., Hollis,
Dorband, \& Yusef-Zadeh 1992; hereafter ME), using programs
that are part of the standard IDL Astronomy
Library\footnote{http://idlastro.gsfc.nasa.gov/}. 

Both ML and ME techniques require a deconvolution ``kernel'' consisting
of the best known representation of the instrumental PSF. 
One then makes an initial estimate of the
``true'' source intensity distribution, i.e., the source intensity
distribution that would be detected 
in the absence of instrumental broadening.
This image is convolved with the PSF and is then compared
with the observed image intensity distribution. The resulting
difference image (or ``residual image'') is stored for reference. The
``true'' image intensity distribution is then modified 
accordingly (following either the ML or ME algorithm prescription), 
and the process is repeated. The change in residual image statistics
(e.g., residual image maximum and standard deviation) over successive 
iterations of either the ML or ME schemes can be used to assess the progress 
of image deconvolution.

\subsection{Reconstructed ISOCAM 4.5 $\mu$m total intensity image}

\begin{figure}[htb]
\includegraphics[scale=0.9,angle=90]{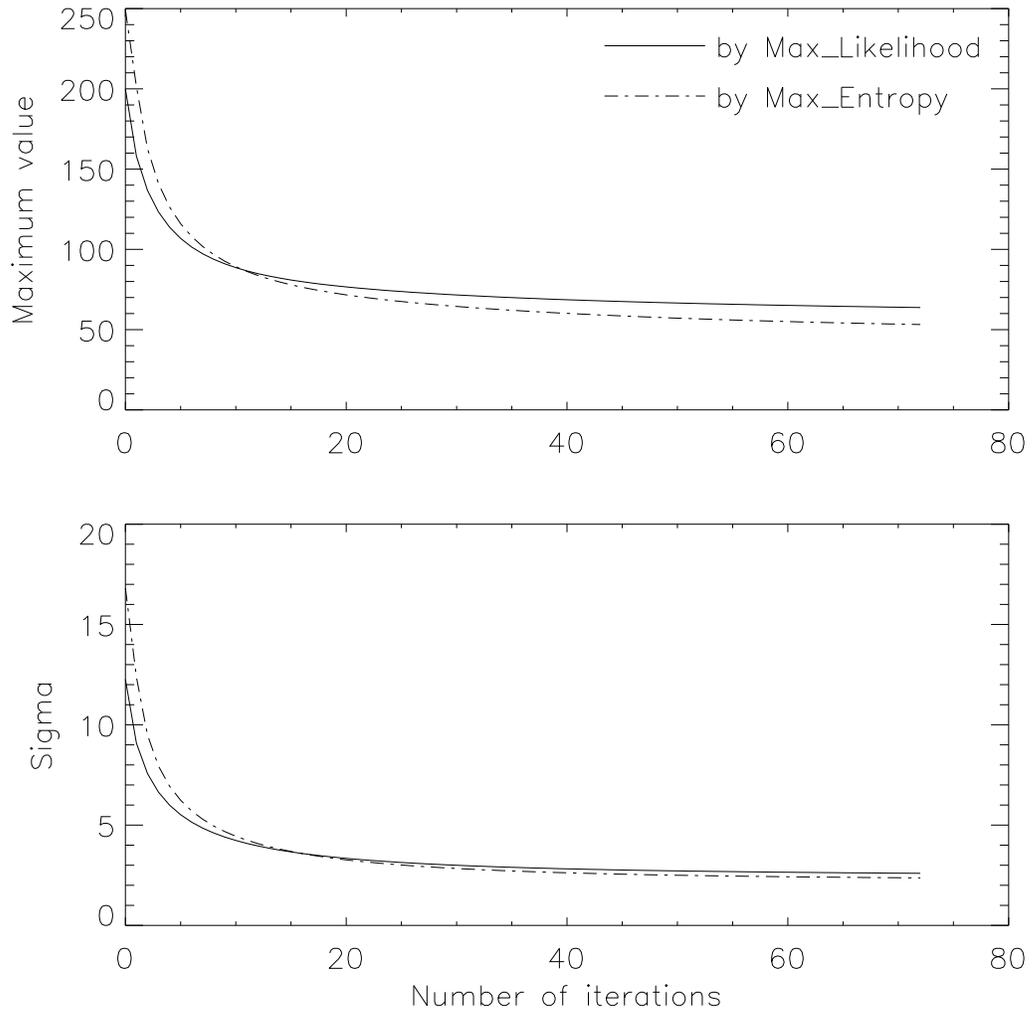}
\caption{ Plot of the maximum value and standard deviation 
of the residual image --- i.e., the difference between the observed
intensity distribution and the inferred, ``true'' intensity
distribution convolved with instrumental PSF ---
as a function of iteration number. Results for maximum likelihood 
and maximum entropy deconvolution techniques are shown as
solid and dashed curves, respectively.}
\label{fig:ML_ME_stats}
\end{figure}

\begin{figure}[htb]
\includegraphics[scale=0.8,angle=90]{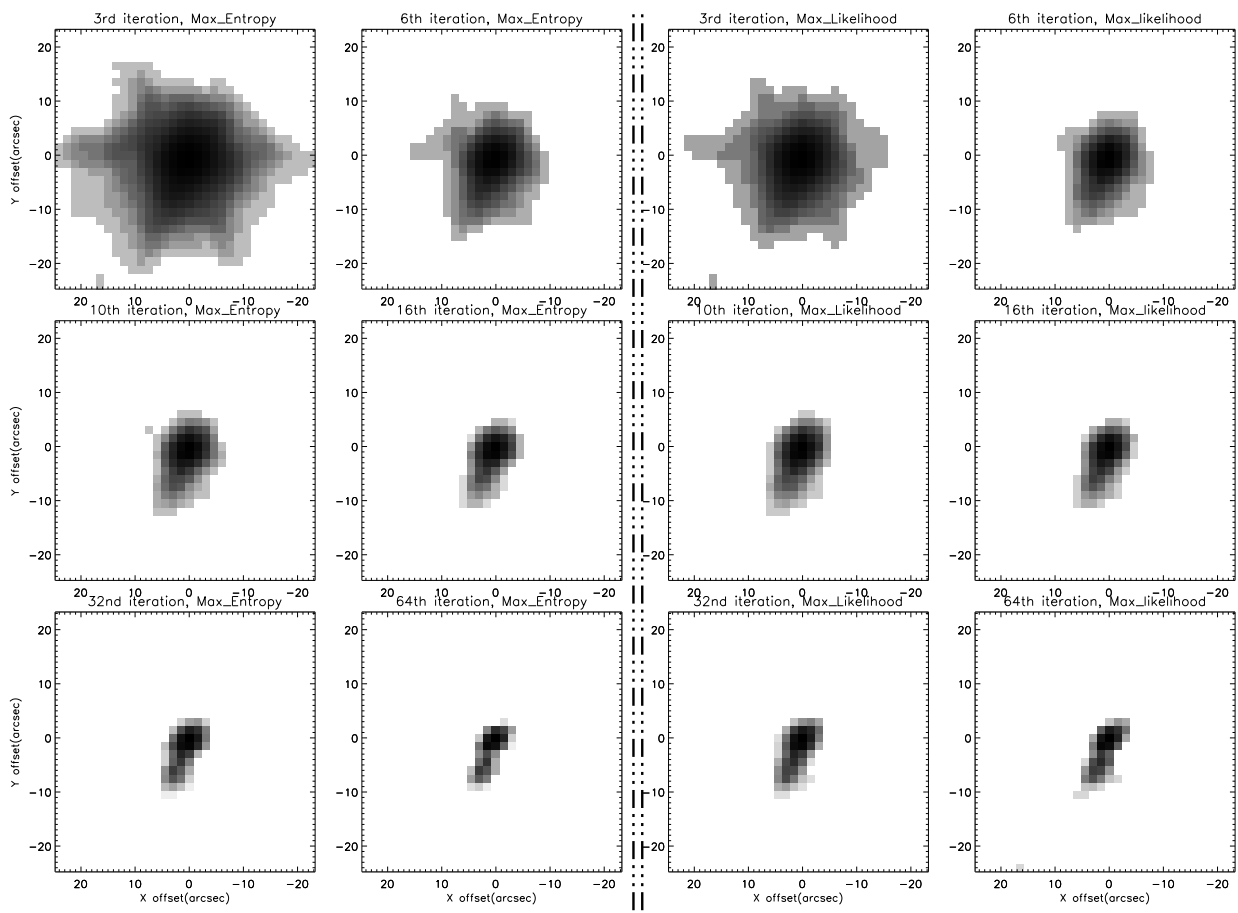}
\caption{ 
Reconstructed total intensity ISOCAM 4.5 $\mu$m images of RAFGL~2688
inferred from the ML (right panels) and ME (left panels) 
techniques. The number of iterations performed to obtain
each reconstructed image increases from upper left to lower right, 
as indicated in each panel.}
\label{fig:ML_ME_images}
\end{figure}

The results of iterative deconvolutions of the ISOCAM LW1 
total intensity image using ML and ME methods 
are displayed in Figs.\ \ref{fig:ML_ME_stats} and \ref{fig:ML_ME_images}. 
For these deconvolutions, we adopted  as the instrumental PSF
the ISOCAM LW1 image of HIC 85317.
It is apparent from both figures that the ML and ME methods produce very 
similar results and, furthermore, that the two methods asymptotically
approach their ``best guesses'' for the ``true'' image at similar
rates. Specifically, after $\sim15$ iterations, the net changes in both 
the maximum and the standard deviation of the residual image
(from their original values) approach $\sim90$\% of the 
differences between their original and asymptotic values (Fig.\ 
\ref{fig:ML_ME_stats}).  We find that, 
for either the ML or ME methods, iterating more than about 
20 times introduces image artifacts, such as negative values and
spurious off-source features. 

Notably, {\it both the ML and ME deconvolution
methods produce the same fundamental result for 
the reconstructed total intensity image} (Fig.\  \ref{fig:ML_ME_images}). 
This reconstructed image consists
of a compact, bright central source, representing 
$\sim$90\% of the total source
intensity, surrounded by fainter nebulosity. Most of the nebulosity
lies to the north-northeast of the bright, compact source. 
This source structure is consistent with the morphology of the
polarimetric map (\S \ref{sec:rawpolmap}), i.e., the bright, central
source appears to illuminate a region of reflection nebulosity 
to its north-northeast.

\subsection{Reconstructed ISOCAM polarization map}

We applied the ML method
to the individual LW1 polarimetric images, to obtain
a deconvolved polarization image set 
$P'_{0}(x,y), P'_{120}(x,y), P'_{240}(x,y)$. For these
deconvolutions, we again adopted the ISOCAM LW1 image of 
HIC 85317 as representative of the instrumental PSF, under the assumption
that the polarizing 
filters produce negligible additional point source distortions. To obtain
each polarimetric image, we
performed 16 iterations of the ML algorithm. The 
resulting deconvolved LW1 polarimetric images of RAFGL~2688 (not shown)
are qualitatively similar to the deconvolved total intensity image
described in the preceding section, i.e., a compact central source
with nebulosity extending to the north-northeast.

The polarization map constructed from the images 
$P'_{0}(x,y), P'_{120}(x,y), P'_{240}(x,y)$ (hereafter referred to as the
reconstructed polarization map) is presented in 
Fig.\ \ref{fig:ML_polmap}. The centrosymmetry evident in the northern
regions of the ``raw'' polarimetric map (Fig.\ \ref{fig:raw_polmap}) is
more pronounced in this reconstructed polarization map. Like the
``raw'' map, the reconstructed
map also displays a minimum of
polarization near the intensity peak in the total intensity image,
and a somewhat disorganized polarization pattern to the south of the
intensity peak.

\section{Discussion}

\subsection{Can the results of the deconvolutions be trusted?}

It is well known that the results of either ML or ME
deconvolution are not necessarily unique. Indeed,  
depending on the quality of both data and PSF model, the results
of either technique can be spurious. Furthermore, we are not aware 
of a rigorous proof that demonstrates that deconvolution of 
polarimetric images preserves polarization information. 
Nonetheless, there are several reasons to conclude that the deconvolution 
results just described are valid. 

\begin{figure}[htb]
\includegraphics[scale=0.9,angle=90]{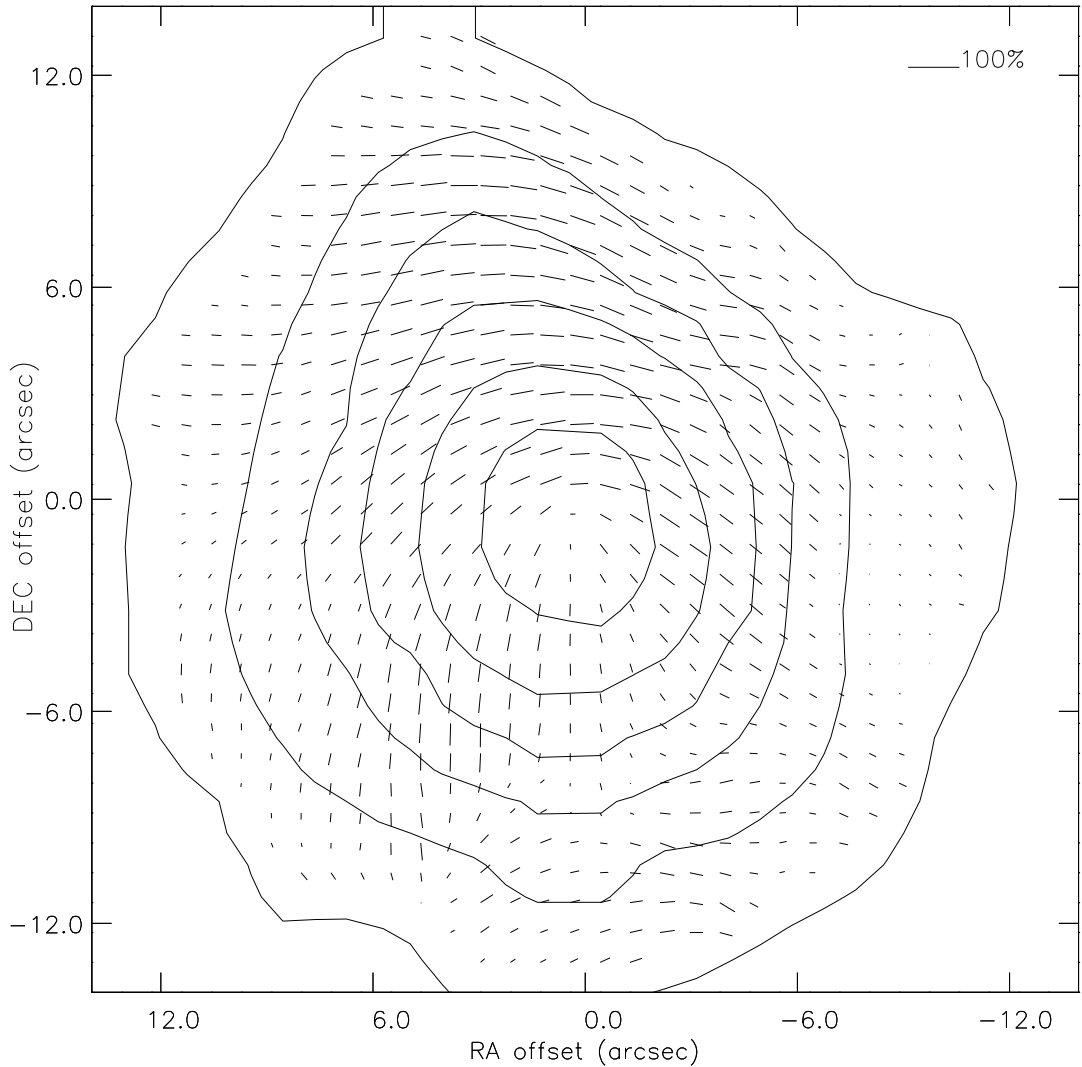}
\caption{ Polarization map of RAFGL~2688
constructed from processed LW1 images obtained through the three
polarizing filters in the LW camera. The orientation and length of 
each vector indicates the direction and relative degree of linear 
polarization, respectively, for a given pixel. Vectors are
displayed for pixels in which the surface brightness is at
least 10 times the pixel-to-pixel
noise level in the individual polarized images.
The map is overlaid on a contour plot
of the total intensity image of Fig.\ 2. Map and
image have been rotated to the celestrial coordinate
reference frame. Contours are displayed at intervals of 0.4
dex, with the lowest contour at 10 times the 
noise level in the individual polarized images.}
\label{fig:raw_polmap}
\end{figure}

First, we find strong consistency between the results
of the ML and ME methods. The two methods converge
to essentially the same end results
(Figs.\ \ref{fig:ML_ME_stats},\ref{fig:ML_ME_images}). 
Second, the reconstructed ISOCAM 4.5 $\mu$m image and polarization
map of RAFGL~2688 are qualitatively consistent with the results of 
previous near-infrared imaging polarimetry and direct mid-infrared
imaging. Specifically, in the reconstructed ISOCAM image, 
RAFGL~2688 is extended along a position angle $\sim15^\circ$, a
result also obtained from high-resolution 2 $\mu$m and 9 $\mu$m imaging 
(Weintraub et al.\ 2000; Morris \& Sahai 2000) but that
can be only marginally inferred from the raw ISO images (e.g., Fig.\
\ref{fig:raw_polmap}). 

\begin{figure}[htb]
\includegraphics[scale=0.9,angle=90]{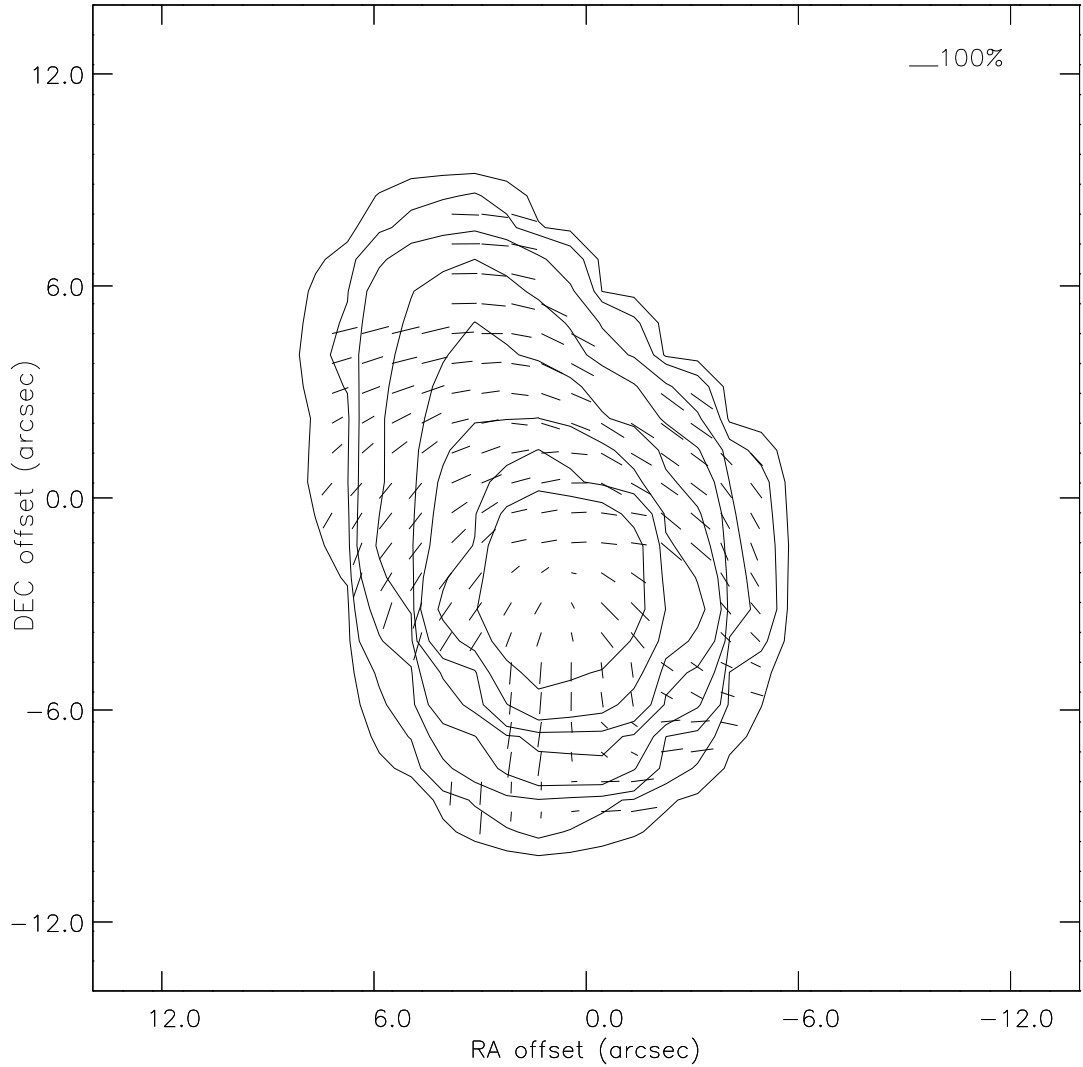}
\caption{ 
Polarization map of RAFGL~2688
constructed from deconvolved LW1 polarimetric images, overlaid on 
a contour plot of the reconstructed total intensity
image. This reconstruction is the result of 16 iterations of
the maximum likelihood method. Polarization vectors are
displayed for pixels in which the surface brightness is at
least 5 times the pixel-to-pixel
noise level in the individual polarized images. Contours are
displayed at intervals of 0.4 dex, with the lowest contour at 5 times the 
noise level in the individual polarized images.}
\label{fig:ML_polmap}
\end{figure}

Third, the reconstructed polarization map 
(Fig.~\ref{fig:ML_polmap}) represents an improvement over the
polarization map constructed from 
``raw'' images (Fig.~\ref{fig:raw_polmap}). In particular, note
the stronger centrosymmetry 
and larger values of polarization in the northern regions of
the map in Fig.~\ref{fig:ML_polmap}. Both improvements would
be expected from a successful deconvolution 
procedure; such a procedure should correct for instrumental
stray light at low spatial frequencies, thereby removing
``contaminating'' flux that would otherwise suppress the
degree of polarization, especially in regions of low
source surface brightness. 

We conclude that
the reconstructed  ISOCAM 4.5 $\mu$m total intensity image
and polarization map of RAFGL~2688 are trustworthy.  

\subsection{The nature of the 4.5 $\mu$m emission from RAFGL~2688} 

The ISOCAM polarimetric map of RAFGL~2688
reveals that the extended nebulosity to the 
north-northeast is highly
polarized, indicative of reflection nebulosity. The
centrosymmetry of this polarization pattern about the 4.5 $\mu$m 
intensity peak indicates that the intensity peak is the likely source 
of illumination of the nebulosity. This source is very
likely a combination of direct photospheric emission from
the central F-type supergiant (e.g., Cohen \& Kuhi 1977) and
photospheric emission reprocessed by dust grains at characteristic
temperatures of $\stackrel{>}{\sim}140$ K (Persi et al.\ 1999).

\begin{figure}[htb]
\includegraphics[scale=0.65,angle=90]{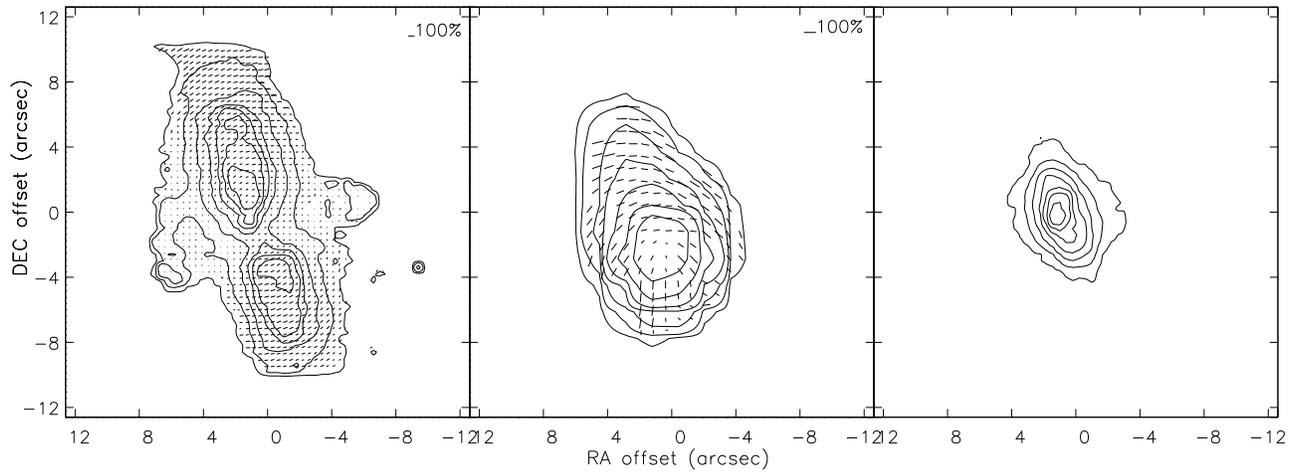}
\caption{ 
Near-infrared polarimetric images of RAFGL~2688
obtained at 2 $\mu$m with HST/NICMOS (left) and at 4.5
$\mu$m with ISOCAM (center), and a direct 8.8 $\mu$m image obtained with
the 10 m Keck telescope and MIRLIN mid-infrared camera 
(right; from Morris \& Sahai 2000).} 
\label{fig:polcomp}
\end{figure}

The apparent ``emergence'' of the central source at longer
infrared wavelengths is seen in
Fig.~\ref{fig:polcomp}. The source is only seen via
reflection at 2 $\mu$m (lefthand panel; see also Weintraub
et al.\ 2000) whereas, evidently, it makes a substantial contribution 
to the nebular morphology (in the form of a strong
point-like source and diluted polarization) near the center
of the nebula at 4.5 and 8.8 $\mu$m (central and righthand panels). Similar
behavior is observed in infrared 
reflection nebulosity surrounding 
young stellar objects, and is usually interpreted as a
decrease in scattering opacity with increasing wavelength
(e.g., Fischer 1995). We caution, however, that the
registration of the three images in
Fig.~\ref{fig:polcomp} is not certain to better than $\sim1''$, 
the ISOCAM polarization imaging suffers from relatively low
spatial resolution, and there is no polarization 
information in the 8.8 $\mu$m image. Furthermore, there are
multiple sources of 2 $\mu$m photons in RAFGL 2688
(Weintraub et al.\ 2000). Hence it may be somewhat premature to 
conclude that the same source that illuminates the nebula at 2
$\mu$m also illuminates reflection nebulosity at 4.5 and 8.8 $\mu$m. 

While it appears we have detected the central illuminating
source at 4.5 $\mu$m, this source evidently is highly obscured along
our line of sight. This conclusion follows from the ratio of
nebular surface brightness to point source intensity; specifically,
at an angular displacement of $6''$ along the polar axis of
the nebula from the central point source, we find that the
surface brightness of scattered light is $\sim$1\% of the
integrated intensity of the central source. For a
reflection nebula whose dust density is inversely proportional
to the square of the angle from the source of
illumination ($\theta$), the scattering
optical depth ($\tau_s$) is related to the ratio of
scattered to incident starlight, $I_s(\theta)/I_0$, via
\begin{equation}
\tau_s(\theta) = 8\pi \; \frac{I_s(\theta)}{I_0} \;
\frac{\theta^2}{\Delta\phi} 
\end{equation}
where $\Delta\phi$ is the solid angle subtended by a pixel
(Jura \& Jacoby 1976; Kastner 1990). Assuming the
integrated intensity of the central source as measured in
our images accurately represents the available
illuminating flux, $I_0$, then the measured ratio
$I_s(6'')/I_0 \approx 0.01$ would suggest a scattering optical
depth $\tau_s(6'') \approx 4$. This result is seemingly
at odds with our detection 
of very high levels of polarization ($\sim60$\%) in this region, however;
such a high degree of polarization requires that
single scattering is dominant over multiple scattering. We
conclude that, along our line of sight, the central object
is highly attenuated 
at 4.5 $\mu$m, such that the measured value of   
$I_s(6'')/I_0$ is an overestimate. A 
conservative lower limit to the degree of 4.5 $\mu$m 
extinction toward the central source is provided by the
requirement that $\tau_s < 1$,
suggesting $A_{4.5\mu m} > 1.5$. 

\begin{figure}[htb]
\includegraphics[scale=0.9,angle=90]{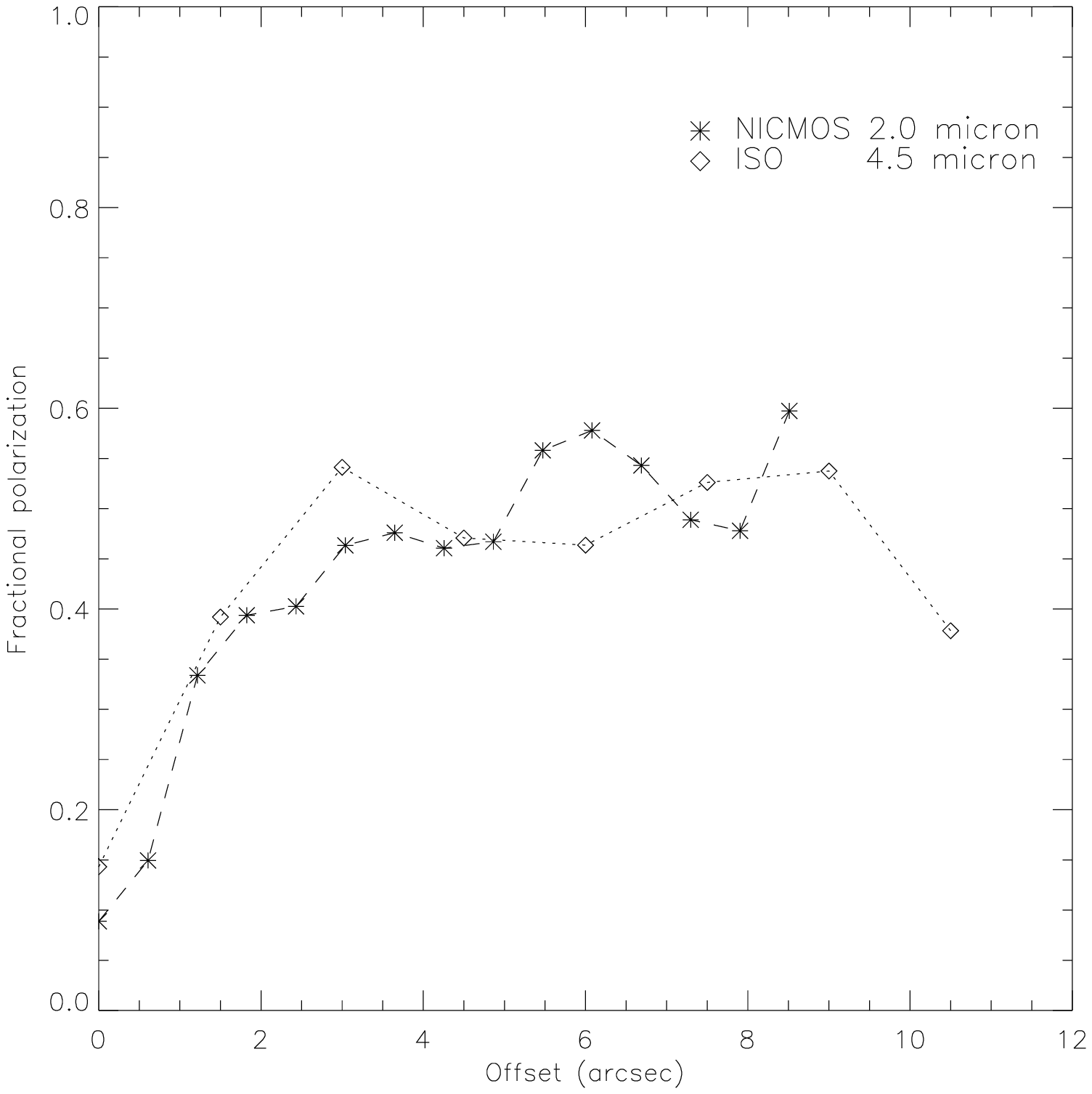}
\caption{ 
Plot of percent polarization vs.\ distance
from the central star, obtained from linear cuts along the
polar axis of RAFGL~2688 in the 2 $\mu$m HST/NICMOS and 4.5 
$\mu$m ISOCAM images.}
\label{fig:polplot}
\end{figure}

Fig.~\ref{fig:polplot} shows a comparison of percent polarization vs.\
radius from the central star, for the 2 $\mu$m (HST/NICMOS) and 4.5
$\mu$m (ISOCAM) polarimetric imaging observations. The figure
demonstrates that percent polarization is more or less independent of
wavelength within the nebula, between $\sim2$ $\mu$m and 5
$\mu$m, at radial offsets well displaced from the star. 
This behavior is similar to that observed in the evolved bipolar
reflection nebula OH 231.8+4.2 (Shure et al.\ 1995),
although the OH 231.8+4.2 observations only extended to 3.6 $\mu$m. From the
large values of polarization in RAFGL~2688 at both 2 $\mu$m and 4.5
$\mu$m, we may safely conclude that scattering 
dominates the nebular emission at these wavelengths; i.e.,
there is little or no contribution from thermal emission
from the dust grains in RAFGL~2688, even at wavelengths as long as $\sim5$
$\mu$m. As no polarization data have been obtained 
for this nebula at longer wavelengths we are, as yet, 
unable to ascertain the nature of the lobe emission seen at
8.8 $\mu$m (Fig.~\ref{fig:polcomp}).

\subsection{Dust grain size distribution}

Most detailed modeling of reflection nebulae at $\sim5$
$\mu$m performed thus far has been geared toward interpeting
large-aperture polarization measurements (e.g., Heckert \&
Smith 1988; Pendleton et al.\ 1990; Fischer 1995; Shure et
al.\ 1995) rather than spatially resolved polarization
imaging. Nevertheless, based on the wavelength 
independence of polarization of RAFGL~2688 and the typical
polarization values of $\sim60$\% at large offsets from the
central star, we
conclude that the bulk of the dust grains in the RAFGL~2688
reflection nebula are significantly smaller than the
wavelengths employed in the HST/NICMOS and
ISOCAM imaging studies.  That is, the grains have typical
radii $a < 1$ $\mu$m. This result, which is similar to that
obtained by Shure et al.\ (1995) for OH 231.8+4.2, follows
from the straightforward analytical treatment of light
scattering by small particles (van de Hulst 1957) as well
as from the more detailed, but limited, modeling of dust
grain populations in infrared reflection nebulae performed
to date (Pendleton et al.\ 1990; Fischer 1995).

\section{Summary}

We have obtained 4.5 $\mu$m polarimetric images of the
evolved bipolar reflection nebula RAFGL~2688 with the CAM
mid-infrared camera aboard the Infrared Space
Observatory. The observations are strongly affected by the
complex CAM point spread function. The source
structure in reconstructed (deconvolved) images is that of a bright
central object with faint 
nebulosity extending to the north-northeast. The position
angle of extended nebulosity is the same as that seen in
ground-based near-infrared images obtained at other wavelengths.
A polarization map constructed from deconvolved,
polarized images displays a large degree of polarization and
a centrosymmetric pattern of position angle within the
faint, extended nebulosity, while the bright central source
displays much lower levels of polarization and disorganized
patterns of polarization position angle. 

The pattern and degree of
polarization within the extended nebulosity, as revealed in
the reconstructed image and polarization map, confirms that
scattering is the dominant source of nebular
light. Comparison with previous space-based (HST/NICMOS)
polarimetric imaging indicates that the degree of
polarization within the extended RAFGL~2688 nebula is
independent of wavelength, suggesting that small ($<1$
$\mu$m) grains do most of the scattering of starlight.

Further, detailed modeling is required to provide 
additional insight into the circumstellar structure and grain
properties of evolved, bipolar nebulae such as RAFGL
2688. It would also be of great value to obtain
mid-infrared imaging polarimetry of RAFGL~2688 at higher
resolution, to ascertain the precise location of the illuminating
source at $\sim5$ $\mu$m. as well as at longer wavelengths, to determine
whether scattering is still important or even dominant beyond 5 $\mu$m.

\acknowledgements{The authors thank Mark Morris and Raghvendra Sahai
for the use of their 8.8 $\mu$m Keck image of RAFGL~2688. CIA is a
joint development by the ESA Astrophysics Division and the ISOCAM
Consortium. The ISOCAM Consortium is led by the ISOCAM PI,
C. Cesarsky. 
ISO research by JHK and JL was supported by JPL grants 1230950 and
1211136 to RIT.}


\begin{references}


\reference {} Blommaert J., Siebenmorgen R., Coulais A., et al., 2001,
``ISO Handbook Volume III (CAM)'', SAI-99-057/Dc, Version 1.2,
{http://www.iso.vilspa.esa.es}

\reference{} Cesarsky C., Abergel A., Agn\`ese P. et al.,
1996, A\&A 315, L32

\reference{} Cohen, M., \& Kuhi, L.V. 1977, ApJ, 213, 79

\reference{} Cox, P., Lucas, R., Huggins, P.J., Forveille, T.,
Bachiller, R., Guilloteau, S., Mallard, J.P., \& Omont, A. 
2000, A\&A, 353, L25

\reference{} Fischer, O. 1995, RMA, 8, 103

\reference{} Heckert, P.A., \& Smith, P.S. 1998, AJ, 95, 873

\reference{} Hollis, J. M., Dorband, J. E., \& Yusef-Zadeh,
F. 1992, ApJ, 386, 293

\reference{} Jones, T.J., \& Dyck, H.M. 1978, ApJ, 220, 159

\reference{} Jura, M., \& Jacoby, G. 1976, Ap\ Letters, 18, 5

\reference{} Jura, M., Turner, J.L., Van Dyk, S., \& Knapp,
G. 2000, ApJ, 528, L105

\reference{} Kastner, J.H. 1990, Ph.D. thesis, UCLA

\reference{} Kastner, J.H., \& Weintraub, D.A. 1995, AJ, 109, 1211

\reference{} Kastner, J.H., Soker, N., \& Rappaport, S.,
eds. 2000a, ``Asymmetrical Planetary Nebulae II: From Origins
to Microstructures,'', ASP Conf.\ Ser.\, Vol.\ 199

\reference{} Kastner, J.H., Weintraub, D.A., Gatley, I., \& Henn, L.
2001, ApJ, 546, 279


\reference{} Kessler M.F., Steinz J.A., Andregg M.E., et al., 1996, A\&A  315,
  L27

\reference{} Lucy, L. B. 1974, AJ, 79, 745

\reference{} Morris, M., \& Sahai, R. 2000, in
``Asymmetrical Planetary Nebulae II: From Origins to
Microstructures,'' eds. J.H. Kastner, N. Soker, \&
S. Rappaport, ASP Conf.\ Ser.\, Vol.\ 199, p.\ 143

\reference{} Ney, E.P., Merrill, K.M., Becklin, E.E., Neugebauer, G.,
\& Wynn-Williams, C.G. 1975, ApJ, 198, L129


\reference {} Ott S., Abergel A., Altieri B., et al., 1996, ASAP
Conference series, Vol. 125, 1997

\reference{} Pendleton, Y., Tielens, A.G.G.M., \& Werner,
M. 1990, ApJ, 349, 107

\reference{} Persi, P., Cesarsky, D., Marenzi, A.R.,
Preite-Martinez, A., Rouan, D., Siebenmorgen, R., Lacombe,
F., Tiphene, D. 1999, A\&A, 351, 201 

\reference{} Richardson, W. H. 1972, OSAJ, 62, 55

\reference{} Sahai, R., Hines, D., Kastner, J.H., Weintraub, D.A., 
Trauger, J. T., Rieke, M. J., Thompson, R. I., Schneider,
G. 1998a, ApJ, 492, L163

\reference{} Sahai, R. et al. 1998b, ApJ, 493, 301

\reference{} Shure, M., Sellgren, K., Jones, T. J., \&
Klebe, D. 1995, AJ, 109, 721

\reference{} Siebenmorgen R., 1996, ``Polarimetric imaging with ISOCAM:
C05 Observer's Manual'', ESA/SAI/96-238/Dc, {http://www.iso.vilspa.esa.es}

\reference{} Siebenmorgen, R.  1999, ``ISO Polarisation Observations,''
R.J. Laureijs and R. Siebenmorgen, ESA-SP435, ISBN 92-9092-740-2,
{http://www.iso.vilspa.esa.es}

\reference{} Siebenmorgen, R., Kr\"ugel, E., 2001, A\&A 364, 625

\reference{} Siebenmorgen, R., Kr\"ugel, E., \& Laureijs, R.J. 2001, A\&A
377, 735

\reference{} Siebenmorgen, R., \& Efstathiou, A. 2001, A\&A 376, L35

\reference{} Tuffs, R.J., Siebenmorgen, R., \& Gallant,
Y.A. 1999, ``Mid-infrared 
polarimetric mapping of the Crab Nebula,'' in: Observing Polarisation
with the Infrared Space Observatory, Eds.: R.J. Lauriejs \&
R. Siebenmorgen, ESA-SPC, available at \\
http://www.iso.vilspa.esa.es/meetings/polarisation/paper/web/

\reference{} van de Hulst, H.C. 1957, ``Scattering of Light
by Small Particles'' (San Francisco: Wiley), p.\ 63

\reference{} Weintraub, D.A., Kastner, J.H., Sahai, R., \& Hines, D. 2000,
ApJ, 531, 401

\reference{} Zuckerman, B., Gilra, D. P., Turner, B. E.,
Morris, M., \& Palmer, P. 1976, ApJ, 205, L15

\end{references}
\end{document}